\documentstyle[11pt,newpasp,twoside,epsf]{article}
\def\ls{{_<\atop^{\sim}}}
\def\gs{{_>\atop^{\sim}}}

\def\edcomment#1{\iffalse\marginpar{\raggedright\sl#1\/}\else\relax\fi}
\reversemarginpar

\begin{document}
\title{The Chandra High Resolution Spectra of Mkn~421}

\author{Fabrizio Nicastro, Antonella Fruscione, Martin Elvis, Aneta 
Siemiginowska}
\affil{Harvard-Smithsonian Center for Astrophysics, Cambridge, MA, USA}
%\author{Antonella Fruscione}
%\affil{Harvard-Smithsonian Center for Astrophysics, Cambridge, MA, USA}
%\author{Martin Elvis}
%\affil{Harvard-Smithsonian Center for Astrophysics, Cambridge, MA, USA}
%\author{Aneta Siemiginowska}
%\affil{Harvard-Smithsonian Center for Astrophysics, Cambridge, MA, USA}
\author{Fabrizio Fiore}
\affil{Osservatorio Astronomico di Monteporzio, Rome, Italy}
\author{Stefano Bianchi}
\affil{Universit\'a degli Studi ``RomaTre'', Rome, Italy}

\begin{abstract}
Mkn~421 was observed by Chandra twice, on November 5, 1999 as part of the 
Chandra calibration program, with the ACIS-HETG configuration, and on 
May 29, 2000 following our Target Of Opportunity request aimed to catch 
the source in an ultra-high state, with both the ACIS-HETG and the HRC-LETG 
configurations. In this contribution we present and compare the two 
{\em Chandra}-MEG observations of Mkn~421, which lasted 26 and 19.6 ks 
respectively. 
\end{abstract}

\section{Introduction}
Mkn~421 (cz = 9234 km s$^{-1}$, Ulrich et al., 1975) is a low redshift 
blazar, intensively studied over all frequencies. 
Recent X-ray observations of this source during a monitoring campaign 
performed by {\em Beppo}-SAX have clearly shown that the beamed continuum 
emission from Mkn~421 undergoes spectral variability correlated 
with flux changes (Malizia et al., 2000, Fossati et al., 2000). 
These spectral changes are driven by the shift of the synchrotron peak 
emission towards higher energies as the source brightens. This causes a 
flattening of the 0.1-10 keV spectrum when the source is in its 
outburst phases. 

%Low resolution X-ray observations of Mkn~421 have never found any strong 
%spectral feature superimposed on the continuum, neither in emission 
%nor in absorption, so confirming the beamed nature of the continuum 
%emission from this source, and the absence of large amount of ionized 
%matter absorbing the beamed continuum along our line of sight. 

\medskip
In this paper we present the first, high-quality, {\em Medium 
Energy Grating} (MEG) high resolution spectra of Mkn~421, taken by 
{\em Chandra} with the ACIS-HETG configuration on November 1999 and 
May 2000. These two observations clearly confirm the 
findings of Malizia et al. (2000) and Fossati et al. (2000), showing a 
marked spectral variability of the nuclear continuum in between them. 
We also look for narrow absorption and emission features that could 
have escaped detection in the previous low-resolution X-ray spectra of this 
source. We present here the results of our analysis. 

\section{Data Reduction and Analysis}
Mkn~421 was observed with the {\em Chandra} ACIS-HETG configuration twice, 
on November 5, 1999 (as a calibration target) and on May 29, 2000, following 
a DDT (Director's Discretionary Time) aimed to catch the source in a very 
high intensity state. 
Both sets of data were processed, and then reduced, with the latest version 
of the {\em Chandra Interactive Analysis of Observations} software (CIAO2.0, 
Elvis, 2001, in preparation), using the most updated calibrations. 
%In particular we used the proper gain-files for each observations (taken 
%with two different temperature of the focal plane CCD detector: -110 C and 
%-120 C, for the November 1999 and the May 2000 observations respectively), 
%the last release of the order-sorting-table files, for the separation of the 
%dispersed orders, and the most up-to-date quantum-efficiency and 
%effective area matrices. 
Standard grade filtering was applied to order-sorted event-2 
files (ASCA grades 0, 2, 3, 4, 6). Spurious events 
accumulated on streaks along the read-out axis of the CCD chips 
during single read-out phases were removed using the 
{\em Chandra} contributed tool {\em destreak} (Houck J.
\footnote{http://asc.harvard.edu/ciao/download\_scripts.html}
). 

First to third order source and background MEG spectra for 
both the observations were extracted from the order-sorted and cleaned 
event-2 files using the tool {\em tgextract} in CIAO2.0. 
%We used default 
%source and background extraction regions, consisting of a 
%rectangular region with width of 50 pixels for the MEG source 
%dispersed spectrum, and two rectangular regions parallel to the source 
%region, above and below it, for the MEG background. 
%The ratio between the background and source region areas is 4.5. 
For each order we rebinned the counts-per-bin histograms by given factors, 
added positive and negative orders, subtracted the background from the 
source spectrum, and divided by the effective area, the exposure time and 
the witdh of the bins in \AA, to end up with binned ``fluxed'' spectra in 
units of ph s$^{-1}$ cm$^{-2}$ \AA$^{-1}$. 

In the following we present the spectral analysis of the MEG first order 
spectra. The compared analysis of all the HETG/LETG data is deferred to a 
forthcoming paper (Nicastro et al., 2001a, in preparation)

\section{Spectral Analysis}

\subsection{The Nuclear Continuum and Its Wavelength-Dependent Variability} 
During the 2000 May Chandra-MEG observation (hereinafter MEG00), Mkn~421 
was more than one order of magnitude brighter than during the previous 
1999 November observation (MEG99), in the entire 2-26 \AA\ band. 
However these changes were wavelength-dependent, as shown, in a 
model-independent way, in Figure 1, where we plot the ratio between the 
MEG00 and MEG99 spectra. 
%the flux, during 
%the high state observation, being almost twice higher in the soft (15-25 
%\AA) than in the hard (2-15 \AA) band when compared with the corresponding 
%fluxes during the 1999 November observation. 
%This is shown in a model-independent way in Figure 1, where we show a 
%histogram of the ratio between the MEG00 and MEG99 spectra. 
%
\begin{figure}
\plotfiddle{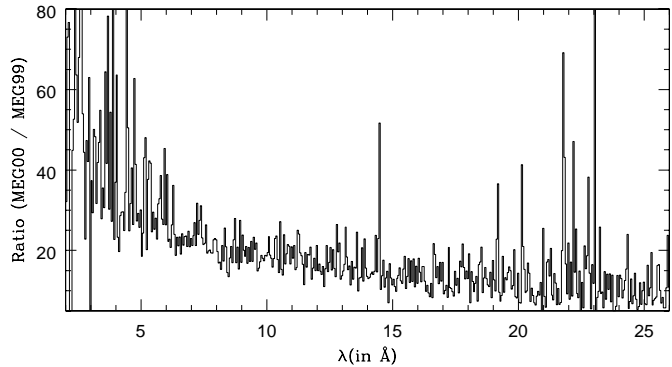}{1.9in}{0}{50}{50}{-150}{-200}
\caption{Ratio between the MEG spectra of Mkn 421 taken during the 
two Chandra observations of 1999 November and the 2000 May.}
\end{figure}
This ratio clearly shows that the source brightened by almost an order of 
magnitude in the soft band (above $\sim 12$ \AA) and by much large factors 
in the hard band ($\lambda \ls 12 \AA$), so flattening considerably between 
the two observations. We note that most of the spikes present in this ratio 
correspond to 1-bin negative fluctuations in the very noise 
MEG99 spectrum (but see also next section). 
%However, some of these spikes, between 18 and 24 \AA, involve more than one 
%adjacent bin and are tentatively identified here as true absorption lines 
%present in the MEG99 spectrum only (see next section). 

Detailed fitting confirms the spectral variability shown by the ratio 
in Figure 1. 
We fitted simultaneously the fluxed MEG spectra of Mkn~421 in {\em Sherpa}, 
using a model consisting of a power law attenuated by interstellar 
absorption (whose hydrogen equivalent column density was constrained to be 
larger than the Galactic column along the line of sight: N$_H = 1.45 
\times 10^{20}$ cm$^{-2}$, Elvis, Lockman \& Wilkes, 1989). 
All the parameters were left free to vary independently between the two 
sets of data. The result of this fit is shown in Table 1 (where power law 
parameters have been converted into energy-fluxed spectra, in units of ph 
$cm^{-2}$ s$^{-1}$ keV$^{-1}$, at 1 keV). Reported errors 
are at a 2-$\sigma$ level for 1 interesting parameter.  
\begin{table}
\caption{Best Fit Continuum Parameters of Mkn 421}
\begin{tabular}{|c|cccc|}
\tableline
Date & $^a$N$_H$ & $\Gamma$ & $^b$Norm & $^c$(0.5-6 keV) Flux \\ 
\tableline
11 Nov 99 & $1.40^{+2.4}_{-0}$ & $2.83^{+0.17}_{-0.05}$ & 
$5.8^{+0.5}_{-0.2}$ & 1.6 \\
29 May 00 & $1.40^{+0.08}_{-0}$ & $2.19 \pm 0.01$ & $93.8 \pm 0.6$ & 33 \\ 
\tableline
\tableline
\end{tabular}
\\$^a$ In $10^{20}$ cm$^{-2}$. 
$^b$ In $10^{-3}$ ph cm$^{2}$ s$^{-1}$ keV$^{-1}$, at 1 keV. 
$^c$ In $10^{-11}$ erg cm$^{-2}$ s$^{-1}$. 
\end{table}
During both the observations the amount of neutral absorption is 
consistent with the Galactic one. The 0.5-6 keV beamed continuum 
of Mkn~421 flattened by $\Delta \Gamma = 0.64^{+0.18}_{-0.07}$ 
between the two observations, while brightening by a factor of $\sim 19$. 

\subsection{High Resolution Spectral Analysis} 
The presence of ionized gas in the 
%reprocessing the nuclear continuum in the 
blazar environment or along the line of sight, between us and the source, 
can be efficiently investigated at high spectral resolution, looking 
for narrow absorption and/or emission features in the soft X-ray band. 
In particular OVII and OVIII ions can imprint strong K$\alpha$ and K$\beta$ 
resonant (R) absorption lines at 21.602 (OVII K$\alpha$), 18.63 (OVII 
K$\beta$) and 18.97 \AA\ (OVIII K$\alpha$) (rest frame), while strong 
intercombination (I) and forbidden (F) OVII K$\alpha$ emission is expected 
at 21.9 and 22.1 \AA\ respectively. 
In Figure 2 we show the 18-24 \AA\ portion of the MEG99 (upper panel) 
and MEG00 (lower panel) spectra of Mkn~421, along with their best fitting 
models (red curves). The spectra are binned by a factor of 10, 
which corresponds to a bin-width of $\Delta \lambda = 0.05$ \AA\ (i.e. R 
= 400 at 20 \AA). 
No clear absorption or emission line is visible in the MEG00 spectrum. 
However, the MEG99 spectrum shows both negative and positive deviations 
from the best fitting continuum model that involve several adjacent bins 
and so are likely to be true, relatively broad, emission and absorption lines. 
We then tried to fit the most significant ($\sim 2 \sigma$ per bin) of 
these features with single positive and/or negative gaussians, with all 
parameters (i.e. normalization, position of the center and full width half 
maximum) free to vary. We fitted up to four gaussians to this portion of 
the spectrum, three in absorption, and one in emission, and tentatively 
identified two different absorbing/emitting systems of ionized gas with 
oxygen mostly distributed between OVII and OIX species. The lines 
identification is based on their relative positions, and we identify a 
system if more than one line from different atomic transitions is detected. 
None of these two systems is seen in the MEG00 spectrum of MKn~421 (Fig. 2, 
lower panel), implying variability of the physical conditions of the 
reprocessing material, and so strongly suggesting its close association 
with the AGN environment. 
At the redshift of Mkn~421 (cz = 9234 km s$^{-1}$, Ulrich M.H. et al., 
1975, ApJ, 198, 261) the resonant OVII K$\alpha$ falls at 22.267 \AA. 
For our two systems we measure, $\lambda_1(OVII K\alpha) = 21.806$ \AA\ 
and $\lambda_2(OVII K\alpha) = 20.153$ \AA. If these identification are 
correct, then the two absorbing/emitting systems are outflowing from the 
central source with velocities of $v_{out}^1 = 6600$, and $v_{out}^2 = 
28500$ km s$^{-1}$ respectively, maybe preceeding the jet along our line 
of sight. If this is the case, the disappearing of these systems during 
the outburst phase of Mkn~421 can be self-consistently accounted for 
by a sharp increase of the ionization degree of the gas due to 
photoionization by the beamed continuum emission. 
\begin{figure}
\plotfiddle{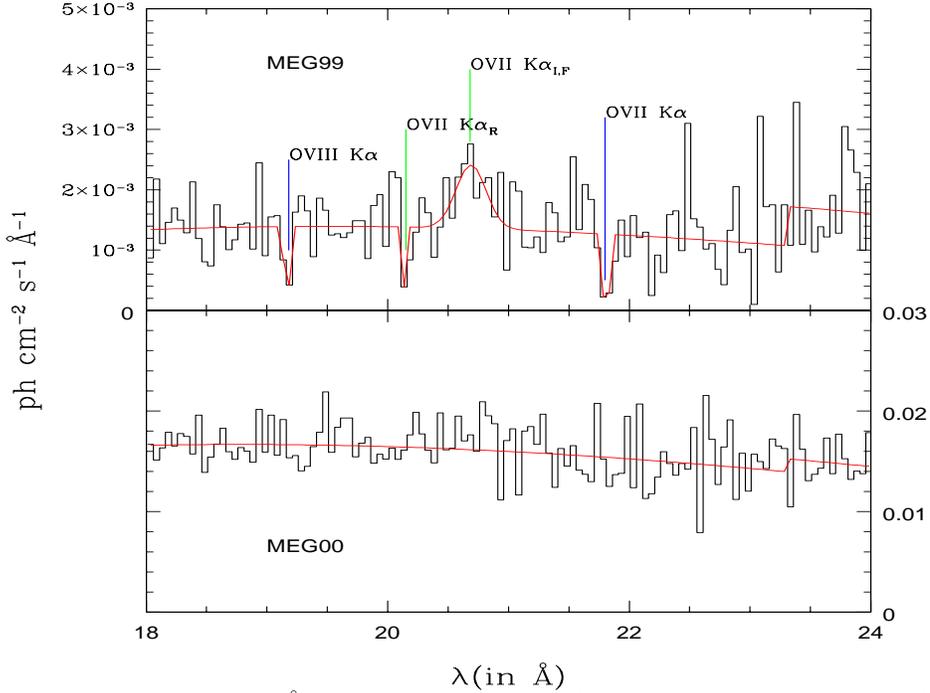}{3.4in}{0}{65}{50}{-210}{-90}
\caption{18-24 \AA MEG99 (upper panel) and MEG00 (lower panel) spectra 
of Mkn~421.}
\end{figure}
\subsection{Searching for the WIGM}
High resolution HST-GHRS observations of Mkn~421 show at least three 
strong absorption lines between 1220 and 1260 \AA\ (Penton, Shull \& Stocke, 
2000, Penton, Stocke \& Shull, 2000). 
%Two of these lines are identified with the SII$\lambda 1250.6$ and 
%SII$\lambda 1253.8$ doublet produced by a known HI High Velocty Cloud in 
%our Galaxy (Lockman and Savage, 1995). 
One of these lines is identified as H Ly$\alpha$ absorption produced by an 
intervening intergalactic system at a redshift of cz = $3035 \pm 6$ km 
s$^{-1}$, close to three galaxies at gas-to-galaxy distances of 2.142, 
3.98 and 4.119 h$_{70}^{-1}$ Mpc. 
The Doppler parameter associated to this line is $b = 35 \pm 5$ km s$^{-1}$, 
and, if the broadening is due only to thermal 
motion of the protons in the gas, it corresponds to a gas temperature of 
$T = (1.4 \pm 0.4 ) \times 10^5 K$, and so to mildly ionized gas in 
thermal equilibrium. This Warm phase of the Intergalactic Medium 
(WIGM) is predicted by hydrodinamical simulation for the formation 
of the large-scale structures in the Universe (e.g. Hellsten et al., 1998; 
Cen \& Ostriker 1999, Dav\'e et al., 2000), and it is thought 
to host the majority of the baryonic matter in the local Universe 
(z$\ls 1$). However, due to the lack of adequate resolution and 
high-contrast spectra in the soft X-ray regime, this fundamental 
baryonic component of the Universe, has so far escaped identification. 

The Doppler parameter for oxygen ions in the intervening system along the 
line of sight to Mkn~421, is 4 times smaller than that measured for 
hydrogen, so suggesting full width half maxima of putative oxygen lines of 
the order of 10 km s$^{-1}$. 
These are far narrower than the resolution of the MEG99 and MEG00 
spectra plotted in Fig. 2, and so, ionized-oxygen absorption associated 
with this intervening system may well have escaped detection at those 
resolution. To detect such a WIGM cloud, a resolution of R $\gs 
1000$ is needed. The full resolution of the MEG at 20 \AA\ is 
of R$\sim 1000$, and an ungroupped MEG spectrum has a bin-width of 
0.005 \AA (i.e. $\sim 70$ km s$^{-1}$ at 22 \AA), which then allows for 
4 non-independent channels for resolution element. At these resolution 
the strongest lines from the WIGM can begin to show-up in spectra with 
adequate signal to noise ratio in the continuum (Nicastro et al., 2001b, 
in preparation). The full resolution MEG00 spectrum of Mkn~421 has $\sim 
10-15$ counts in the continuum at $\sim 20$ \AA, and so can be used to 
investigate the presence of such features. 

Based on the observed properties of the UV Ly$\alpha$ system, 
we ran our resonant absorption models (Nicastro, Fiore and Matt, 1999) 
for a cloud of collisionally ionized gas with a temperature of $T = 1.4 
\times 10^5$ K, to predict what lines we should expect in the MEG00 
spectrum of Mkn~421, and at which sensitivity level. 
We included the photoionization contribution by the diffuse EUV-to-X-ray 
background at the redshift of the intervening Ly$\alpha$ system of Mkn~421 
(Nicastro et al., 2001b). The relative contribution of photoionization 
depends, of course, on the baryonic density in the cloud. 
At the redshift of this cloud the photoionization contribution becomes 
important for $n_b \ls 10^{-5}$ cm$^{-3}$. At higher density the dominating 
mechanism is collisional ionization and the gas is only mildly ionized with 
oxygen distributed between OIV and OVI. At these densities no line is 
expected to show up in high resolution X-ray spectra. For gas density of 
$n_b \sim 10^{-6}$ cm$^{-3}$, OVII has its maximum, and strong OVII 
K$\alpha$ and K$\beta$ are expected at $\lambda = 21.821$ \AA\ and 
$\lambda = 18.819 \AA$. Finally for densities lower than $n_b \sim 10^{-7}$ 
photoionization dominates and the gas is almost fully ionized, with the 
oxygen mainly in its OVIII-OIX stages. In this case resonant OVIII 
K$\alpha$ at 19.162 \AA\ is predicted. 

We then examined the full resolution MEG00 spectrum looking for these 
features. We did not find any significant absorption line at the expected 
wavelengths, and put the following $1-\sigma$ upper limits on the observed 
equivalent widths: EW(OVIIK$\alpha$) $> -15$ m\AA, EW(OVIIK$\beta$) $> -3$ 
m\AA, and EW(OVIIIK$\alpha$) $> -3$ m\AA. These are still compatible with 
a WIGM filament with typical metallicities of $Z = 0.01-0.1 Z_{\odot}$, 
equivalent hydrogen column densities of $N_H = 10^{18}-10^{20}$ cm$^{-2}$, 
and densities lower than $10^{-5}$ cm$^{-3}$. Higher signal to noise 
observations are then needed to definitively address this important issue 
and constrain the volume and column densities of the local Ly$\alpha$ 
filament along the line of sight to Mkn 421, which in turn would allow 
one to estimate the baryonic mass contained in this filament. 

\section{Summary}
In this paper we presented the first high resolution X-ray 
spectra of Mkn~421, taken with the {\em Chandra}-HETG on 
November 5, 1999 and May 29, 2000. 
The main results of our analysis can be summarized as follows: 
(a) the 0.5-6 keV continuum of Mkn~421 flattened by $\Delta \Gamma = 
0.64^{+0.18}_{-0.07}$ between the two {\em Chandra} observations, while 
brightening by a factor of $\sim 19$. (b) Two absorbing/emitting systems 
of highly ionized gas have been tentatively identified in the low-state 
MEG spectrum of Mkn~421 through the detection of absorption and emission 
K$\alpha$ lines from OVII and OVIII ions. These two systems are
not seen in the high-state spectrum of Mkn~421, so suggesting that they 
were intrinsic to the nuclear environment and became fully ionized, 
and so transparent, as the source brightened. If this interpretation 
is correct, the two systems were both outflowing from the central source 
with velocities of 6600 and 28500 km s$^{-1}$, maybe preceeding the jet 
along our line of sight. 
(c) Finally we demonstrated that both very high spectral resolution 
and high signal to noise data are required to detect high-ionization 
absorption lines from the WIGM, and used the high signal to noise MEG 
spectrum of Mkn~421 at full resolution to put limits on the equivalent 
widths of the predicted OVII and OVIII K$\alpha$ and K$\beta$ resonance 
absorption lines that are likely to be associated to the intervening H 
Ly$\alpha$ system along the line of sight to this source.

\end{document}